\documentclass[12pt]{article}
\usepackage{epsf, cite, amssymb}
\usepackage{epsfig}
\usepackage{amsmath,amssymb}
\usepackage{latexsym}
\usepackage{epsfig}
\usepackage{cite}
\usepackage[english]{babel}
\usepackage{graphicx,color}

\makeatletter
\renewcommand\section{\@startsection {section}{1}{\z@}%
                               {-3.5ex \@plus -1ex \@minus -.2ex}
                               {2.3ex \@plus.2ex}%
                               {\normalfont\large\bfseries}}
\renewcommand\subsection{\@startsection{subsection}{2}{\z@}%
                                 {-3.25ex\@plus -1ex \@minus -.2ex}%
                                 {1.5ex \@plus .2ex}%
                                 {\normalfont\bfseries}}
\makeatother

\def\IZ{\relax\ifmmode\mathchoice
{\hbox{\cmss Z\kern-.4em Z}}{\hbox{\cmss Z\kern-.4em Z}}
{\lower.9pt\hbox{\cmsss Z\kern-.4em Z}} {\lower1.2pt\hbox{\cmsss
Z\kern-.4em Z}}\else{\cmss Z\kern-.4em Z}\fi}
\def\IR{\relax{\rm I\kern-.18em R}}

\def\one{{\hbox{ 1\kern-.8mm l}}}

\def\tr{{\rm tr\,}}

\newlength{\bredde}
\def\slash#1{\settowidth{\bredde}{$#1$}\ifmmode\,\raisebox{.15ex}{/}
\hspace*{-\bredde} #1\else$\,\raisebox{.15ex}{/}\hspace*{-\bredde}
#1$\fi}

\newsavebox{\zzzbar}
\sbox{\zzzbar}
{\setlength{\unitlength}{0.9em}
\begin{picture}(0.6,0.7)
\thinlines
\put(0,0){\line(1,0){0.6}}
\put(0,0.75){\line(1,0){0.575}}
\multiput(0,0)(0.0125,0.025){30}{\rule{0.3pt}{0.3pt}}
\multiput(0.2,0)(0.0125,0.025){30}{\rule{0.3pt}{0.3pt}}
\put(0,0.75){\line(0,-1){0.15}}
\put(0.015,0.75){\line(0,-1){0.1}}
\put(0.03,0.75){\line(0,-1){0.075}}
\put(0.045,0.75){\line(0,-1){0.05}}
\put(0.05,0.75){\line(0,-1){0.025}}
\put(0.6,0){\line(0,1){0.15}}
\put(0.585,0){\line(0,1){0.1}}
\put(0.57,0){\line(0,1){0.075}}
\put(0.555,0){\line(0,1){0.05}}
\put(0.55,0){\line(0,1){0.025}}
\end{picture}}

\newcommand{\ena}{\end{eqnarray}}
\newcommand{\beqa}{\begin{eqnarray}}
\newcommand{\eeqa}{\end{eqnarray}}
\newcommand{\bea}{\begin{eqnarray}}
\newcommand{\eea}{\end{eqnarray}}

\newcommand{\be}{\begin{equation}}
\newcommand{\ee}{\end{equation}}

\usepackage{graphicx}

\def\be{\begin{equation}}
\def\ee{\end{equation}}
\def\beq{\begin{eqnarray}}
\def\eeq{\end{eqnarray}}

\def\({\left (}
\def\){\right )}
\def\[{\left [}
\def\[{\right ]}

\def\tr{\mathrm{tr}}

\def\ba{\begin{eqnarray}}
\def\ea{\end{eqnarray}}

\input amssym.def
\input amssym.tex

\newcommand{\bbibitem}[1]{\bibitem{#1}\marginpar{#1}}
\def\Bibitem#1{\bibitem{#1}%
  \smash{\hbox to0pt{\raise1ex\hbox{\tiny[#1]}\hss}}}

\def\Label#1{\label{#1}%
  \smash{\hbox to0pt{\raise1ex\hbox{\tiny[#1]}\hss}}}
\def\noLabels{\let\Label=\label}
\def\nobbibitem{\let\bbibitem=\bibitem}
 \def\noBibitem{\let\Bibitem=\bibitem}

\begin{document}

\begin{titlepage}
\begin{flushright}
arXiv:
\end{flushright}
\vfill
\begin{center}
{\Large \bf Thermalization of mutual and tripartite information in strongly coupled two dimensional conformal field theories}

\vskip 10mm

{\large V.~Balasubramanian$^{a}$, A.~Bernamonti$^{b}$, N.~Copland$^{b,c}$,\\
\vspace{3mm}
 B.~Craps$^b$ and F.~Galli$^{b}$}

\vskip 7mm

$^a$ David Rittenhouse Laboratory, University of Pennsylvania, \\
\hspace*{0.15cm} Philadelphia, PA 19104, USA. \\
$^b$ Theoretische Natuurkunde, Vrije Universiteit Brussel, \\
\hspace*{0.15cm}  and International Solvay Institutes, \\
\hspace*{0.15cm} Pleinlaan 2, B-1050 Brussels, Belgium. \\
$^c$ Centre for Quantum Spacetime, Sogang University,\\
\hspace*{0.15cm}  Seoul 121-742, Korea.\\

\vskip 3mm 
\vskip 3mm 
{\small\noindent  {\tt vijay@physics.upenn.edu, Alice.Bernamonti@vub.ac.be,  ncopland@sogang.ac.kr, \\
Ben.Craps@vub.ac.be, Federico.Galli@vub.ac.be}}

\end{center}
\vfill

\begin{center}
{\bf ABSTRACT}
\vspace{3mm}
\end{center}
The mutual and tripartite information between pairs and triples of disjoint  regions in a quantum field theory
 are sensitive probes of the spread of correlations in an equilibrating system.
We compute these quantities in strongly coupled two-dimensional conformal field theories with a gravity dual following the homogenous deposition of energy.    The injected energy is modeled in anti-de Sitter space as an infalling shell, and the information shared by disjoint intervals is computed in terms of geodesic lengths in this background.   For given widths and separation of the intervals, the mutual information typically starts at its vacuum value, then increases in time to reach a maximum, and then declines to the value at thermal equilibrium.      A simple causality argument qualitatively explains this behavior.   The tripartite information is generically non-zero and time-dependent throughout the process.    This contrasts with (but does not contradict) the time-independent tripartite information one finds after a two-dimensional quantum quench in the limit of large time and distance scales compared to the initial inverse mass gap. 
\vfill


\end{titlepage}


\section{Introduction}\label{intro}

Consider a quantum field theory whose state is described by the density matrix $\rho$.   Measurements made within a spatial region $A$ (with complement $\bar{A}$) can all be described in terms of the reduced density matrix $\rho_A = \tr_{\bar{A}} \rho$.   The entanglement entropy of $A$ is defined as $S(A) = -\tr \left[ \rho_A \log \rho_A \right]$.  If the full system is in a pure state, $S_A$ equals $S_{\bar{A}}$ and measures the amount of entanglement between $A$ and $\bar{A}$.     Now consider two disjoint regions $A$ and $B$.   The amount of correlation between these regions (both classical and quantum) is measured by the {\it mutual information}
\be
I(A,B) =  S(A) + S(B) - S(A \cup B) \, .
\label{MIform}
\ee
While the entanglement entropy of a spatial region $A$ has a UV divergence proportional to the area of the boundary of $A$, the mutual information is finite, making it an especially convenient quantity to study.   The mutual information appears in bounds that limit how well an observer of $A$ can predict events in $B$.   Indeed $I(A,B) \geq 0$ with equality if and only if $A$ and $B$ are uncorrelated, i.e. $\rho_{A \cup B} = \rho_A \otimes \rho_B$.   Mutual information can potentially provide a powerful description of how correlations evolve and spread in an out-of-equilibrium system.

Progress in computing mutual information has been made for certain two-dimensional conformal field theories in special limits \cite{Calabrese:2010he}.    In particular, Calabrese and Cardy have studied the evolution of entanglement entropy following a quantum quench \cite{Calabrese:2005in}.   They studied quenches in which a two-dimensional field theory with a gap suddenly becomes conformal (e.g., because a parameter in the Hamiltonian such as an external field is changed).  The evolution of mutual information in this system can be computed in limits where the time, the interval sizes, and the interval separations are all large compared to the initial inverse mass gap.   It was found that the entanglement entropy of an arbitrary number of intervals (and hence the mutual information defined above) exhibits a piecewise linear behavior in time.   This behavior is consistent with a simple model based on causal propagation with the speed of light from an initial state with short-range correlations.

Another interesting quantity to consider in this context is the {\it tripartite information}, which measures the degree of extensivity of the mutual information. This quantity is defined for three spatial  regions $A$, $B$ and $C$ as
\bea
I_{3}(A,B,C)& =&  S(A) + S(B) + S(C) - S(A \cup B)-S(A \cup C)-S(B \cup C) + S(A \cup B \cup C)  \nonumber \\
&=& I(A,B) + I(A,C) - I(A, B\cup C)\, . 
\label{TInform}
\eea
By definition, it is symmetric under permutations of its arguments. 
In a generic field theory, depending on the choice of the regions,  it can be  positive, negative or zero, meaning that the mutual information is subextensive, superextensive or extensive, respectively. It does not have the UV divergences present in the entanglement entropy. This is true even when the three regions share boundaries (in contrast to the mutual information which is cut-off dependent in this case and needs to be renormalized). 

For equilibrium states of field theories that have a gravity dual, the AdS/CFT correspondence gives us a simple recipe for computing entanglement entropy in terms of the areas of minimal surfaces in a dual spacetime \cite{Ryu:2006bv}.   The mutual and the tripartite information between regions can then be computed through \eqref{MIform} and \eqref{TInform}.    Applying this method, an interesting phase transition was observed, where the mutual information between a pair of disjoint intervals in two-dimensional field theory  is of $O(1)$ for large separations, but of order the central charge for small separations \cite{Ryu:2006bv, Headrick:2010zt}.  A recent work  \cite{Hayden:2011ag}  has also proven that for quantum field theories at equilibrium with a holographic dual, the mutual information is always extensive or superextensive, meaning that $I_{3}(A,B,C) \le0$.

 The purpose of our paper is to understand how these results extend to a dynamical setting where energy is injected at $t=0$ and then proceeds to equilibrate.  We model the injection of energy as a shell of null dust falling into anti-de Sitter (AdS) space.  We then apply a proposal of \cite{Hubeny:2007xt} for computing entanglement entropy in a time-dependent AdS/CFT setting in terms of the areas of extremal surface in the dual spacetime.   A key difference between our setting and that of \cite{Calabrese:2005in} is that our theory is always conformal and hence starts with long range correlations, whereas their model is gapped at early times so that the correlations are essentially local \cite{AbajoArrastia:2010yt}.  

We find that the mutual information between intervals in our setting starts at the vacuum value and ends at the thermal value, but usually passes through an intermediate phase where it is higher than either.  The rise and fall are nearly linear in time.  A causality argument (slightly modified from \cite{Calabrese:2005in,AbajoArrastia:2010yt}) explains qualitative features of our results. We find that the tripartite information is non-positive in all the cases that we study, matching the expectation based on the results of \cite{Hayden:2011ag} who studied static configurations.  In our dynamical setting the tripartite information changes with time during the process of thermalization in many parameter regimes.   In contrast, the Calabrese and Cardy results for two-dimensional quantum quenches, where a mass gap is suddenly taken to zero, imply that $I_3$ is constant in their setting \cite{Calabrese:2005in}.    If the AdS/CFT entanglement entropy proposal of  \cite{Hubeny:2007xt} is correct, the difference is probably due to the fact that our initial state has long distance correlations. In other words, the ``mass gap" in our setup would be zero, so that we are studying the system on time and distance scales small compared to the would-be inverse initial mass gap.


\section{Mutual information}\label{sec:MI}

\begin{figure}[htbp]
\begin{center}
\includegraphics[width=0.4 \textwidth]{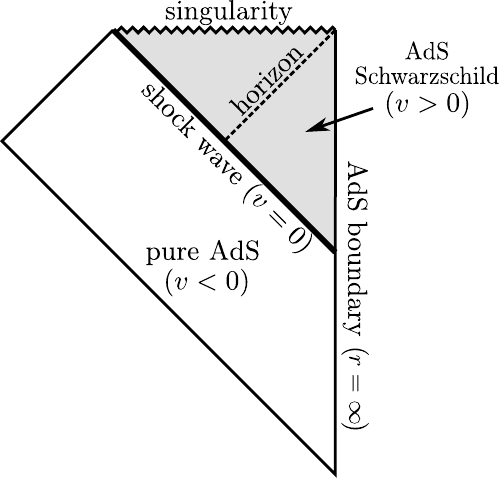}
\hfil 
\includegraphics[width=0.5\linewidth]{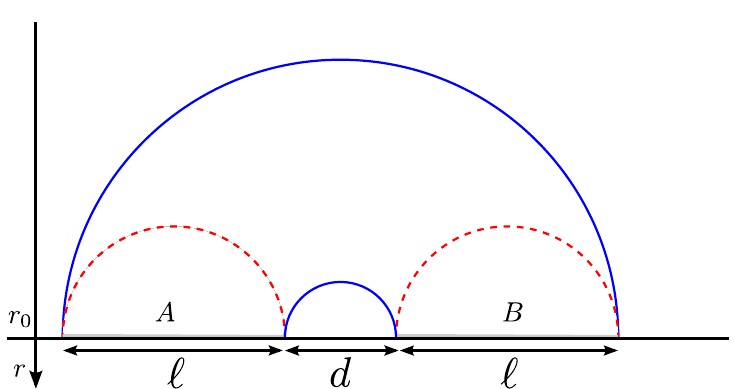} \\ 
({\bf a}) \hfil ({\bf b}) \\
\end{center}
\caption{({\bf a}) The causal structure of the Vaidya spacetime shown in the Poincar\'{e} patch of AdS space.  The asymptotic boundary (vertical line on the right hand side) is planar, and the null lines on the left hand side of the diagram represent the Poincar\'{e} horizon.  ({\bf b}) Connected (in blue) and disconnected (in red dashed) locally minimal surfaces for the boundary region $A \cup B$ in AdS$_3$. 
} \label{fig:VaidyaSpacetime}
\end{figure}

Consider the 3-dimensional infalling  shell geometry described by the Vaidya metric
\be
ds^2 =- \left[r^2 - r_H^2 \Theta(v) \right]dv^2 + 2 dr\, dv +r^2 dx^2 \,,
\label{eq:Vaidya}
\ee
where $v$ labels ingoing null trajectories, $\Theta$ is the step function and we have set the AdS radius equal to 1.
In this form, the Vaidya metric describes a zero thickness shell composed of tensionless null dust (Fig.~\ref{fig:VaidyaSpacetime}a).
Outside the infalling shell ($v>0$), the geometry is identical to that of an AdS black brane with Hawking temperature $T=r_H/2\pi$, while inside the shell ($v<0$) it is the same as that of pure
AdS.  
The change of coordinates
\bea
v = \left\{\begin{array}{ll} t + \frac{1}{2r_H} \ln  \frac{|r-r_H|}{r+ r_H}\, & \quad v>0 \\
                                            t-\frac 1 r\, & \quad v<0 
                                             \end{array}\right.
\eea
brings the two metrics to the standard form
\bea
ds^2 = \left\{\begin{array}{ll} - (r^2 - r_H^2)dt^2 + \frac{dr^2}{r^2 - r_H^2} + r^2 dx^2\, & \quad  v>0 \\
                                            - r^2 dt^2 + \frac{dr^2}{r^2}+r^2 dx^2\, & \quad v<0 
                                             \end{array}\right.
\eea
On the boundary, at $r = \infty$, the coordinates $v$ and $t$ coincide. 

We consider the time evolution of the  mutual information between disjoint intervals of length $\ell$ separated by a distance $d$ on the boundary of the Vaidya spacetime \eqref{eq:Vaidya}.   In three bulk dimensions the entanglement entropy of a connected region $A$ on the spacetime boundary is proposed to be computed holographically by the length of the bulk geodesic that connects the endpoints of $A$ \cite{Hubeny:2007xt}.   For disjoint intervals $A$ and $B$ there are multiple candidates for the geodesics that contribute to $S(A \cup B)$.
 In \cite{Headrick:2010zt}, it was shown that for equilibrium configurations, $S(A \cup B)$ is given by the minimum of the lengths of two sets of geodesics (shown as solid and dashed lines in Fig.~\ref{fig:VaidyaSpacetime}b).  We will assume that this prescription extends to the dynamical Vaidya background  as a rule saying that $S(A\cup B)$ is determined by the length of the shortest collection of geodesics connecting the endpoints of $A$ and $B$.\footnote{There is also a topological condition \cite{Headrick:2010zt} that the geodesics should be continuously deformable to the AdS boundary.  This condition will not come into play for us.}  

The geodesics connecting two equal-time boundary endpoints $(t_0, 0)$ and $(t_0,\ell)$ have been studied in \cite{AbajoArrastia:2010yt,Balasubramanian:2010ce}.  These results have been extended to non-equal time spacelike geodesics in \cite{Aparicio:2011zy}.   Since the geodesic length ${\cal L}$ diverges due to contributions near the AdS boundary, it is more convenient to work with a regularized length computed by introducing a large radius cutoff in AdS at $r=r_0$ and defining  $\delta {\cal L} \equiv {\cal L} - 2 \ln(2 r_0)$.  This removes the divergent part of the length in the vacuum state (pure AdS).  For $t_0 \le 0$ (before the shell falls in) the geodesics lie entirely in AdS at fixed time $t= t_0$. Their renormalized length is
\be
 \delta {\cal L}_{\text{vacuum}}(\ell)
 =2\ln{\frac \ell 2}\,.
 \label{LvsellAdSren}
\ee
For $0 < t_0 < \ell/2$, the geodesics cross the infalling shell and their renormalized length is
\begin{align}
\label{Lren}
\delta {\cal L}_{\text{shell}}( \ell, t_0)
& = 2 \ln \left[ \frac{\sinh(r_H t_0)}{r_H s(\ell, t_0)} \right]\,,
\end{align}
where $s(\ell, t_0) \in [0,1]$ is parametrically defined by
\begin{align} \label{seq}
 \ell
 =
 {1\over r_H}\left[{2c\over s\rho}+\ln\left({2(1+c)\rho^2+2s\rho-c \over 2(1+c)\rho^2-2s\rho-c}\right)\right]\,,
\end{align}
with $c = \sqrt{1-s^2}$ and 
\begin{equation}
2 \rho = \coth (r_H t_0) + \sqrt{\coth^2(r_H t_0) -\frac{2c}{c+1}}\,.
\end{equation}
For $t_0 \ge \ell/2$, the geodesics lie on the  $t=t_0$ surface in the black brane geometry and 
\begin{align}
 \delta {\cal L}_{\text{thermal}}(\ell)
 =2\ln{\sinh{r_H\ell\over 2}\over r_H}\,.
 \label{LvsellBTZren}
\end{align}
In terms of these geodesic lengths our formula for the mutual information between the intervals $A$ and $B$ is
\begin{equation} \label{mutualinfo}
 I({A,B}) = \frac{1}{4 G_N} \left\{ 2 \delta {\cal L}(\ell) - \text{Min}\left[ 2 \delta {\cal L}(\ell),  \delta {\cal L}(2\ell +  d) +  \delta {\cal L}(d)\right] \right\} \, .
\end{equation}
Observe that in principle we should also consider geodesics of spatial separation $\ell + d$, which connect the two left (right) end-points of the two intervals A and B. 
We should thus evaluate in \eqref{mutualinfo}
\be \label{eq:Min}
 \text{Min}\left[ 2 \delta {\cal L}(\ell),  \delta {\cal L}(2\ell +  d) +  \delta {\cal L}(d), 2 \delta {\cal L}(\ell +d ) \right] \,.
\ee
However, for any fixed boundary time $t_0$:
\bea \label{lengthineq}
\delta {\cal L}_{\text{vacuum}} (\ell + d) &\ge& \delta {\cal L}_{\text{vacuum}} (\ell)\, , \nonumber \\ 
\delta {\cal L}_{\text{thermal}} (\ell + d) &\ge& \delta {\cal L}_{\text{thermal}} (\ell)\,  , \\ 
 \delta {\cal L}_{\text{shell}} (\ell + d, t_0) &\ge& \delta {\cal L}_{\text{shell}} (\ell, t_0)\,, \nonumber 
\eea
since, in each case, $\delta {\cal L}$ is monotonically increasing as a function of  the spatial boundary separation, as is seen from \eqref{LvsellAdSren}, \eqref{LvsellBTZren} and numerically from \eqref{Lren}.  Therefore, when all the possible competing geodesics are of the same type (in terms of crossing the shell or not),  it is enough to consider \eqref{mutualinfo}. There are however time intervals in which the property \eqref{lengthineq} is not enough to exclude the case of intersecting geodesics of length $ \delta {\cal L}(\ell +d )$ from \eqref{eq:Min}. For example, when $\ell > d$ and $t_0 \in [\ell/2 , (\ell + d)/2)$,  one has  $\delta {\cal L} (\ell + d) = \delta {\cal L}_{\text{shell}} (\ell+d, t_0)$ while $\delta {\cal L} (\ell ) = \delta {\cal L}_{\text{thermal}} (\ell)$. In these cases, however, it follows from \eqref{Lren} and \eqref{LvsellBTZren} that
\be
 \delta {\cal L}_{\text{shell}} (\ell + d, t_0) \ge \delta {\cal L}_{\text{thermal}} (\ell)
\ee
in the relevant time interval, and it is therefore always enough to consider \eqref{mutualinfo}.

\begin{figure}[t]
\centering
\includegraphics[width=0.5\linewidth]{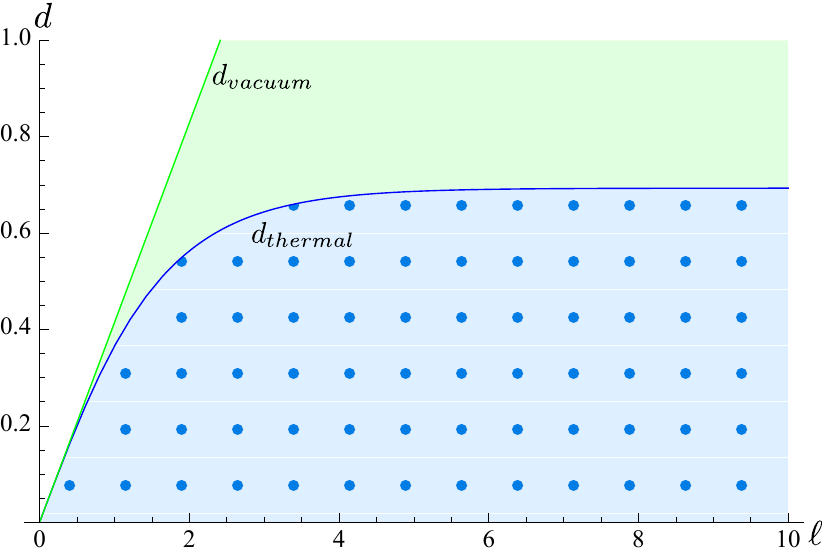}
\caption{$d_{\text{thermal}}$ and $d_{\text{vacuum}}$ as a function of $\ell$ for $r_H=1$. The mutual information surely vanishes for $t_0 \le 0$ and $t_0 \ge (2\ell+d)/2$ for $d$ and $\ell$ in the white region, for  $t_0 \ge (2\ell+d)/2$ in the green shaded region, while it is generically everywhere non-zero in the blue dotted region. 
}
\label{fig:dlAdS}
\end{figure}

In the Vaidya spacetime \eqref{eq:Vaidya}, for $t_0 \le 0$, the mutual information of  $A$ and $B$  coincides with the vacuum AdS result of  \cite{Headrick:2010zt}:
\be
I(A,B)_{\text{vacuum}} = 
\left\{\begin{array}{cc}0 \,,& d \ge d_{\text{vacuum}}  \\
     					  \frac c 3 \ln \left[\frac{\ell^2}{d(2\ell+d)}\right] \,,& d \le d_{\text{vacuum}}
					 \end{array}\right.
\label{MIAdS}
\ee			 
where we have used \eqref{LvsellAdSren}, the central charge $c = 3/(2 G_N)$ and $d_{\text{vacuum}}(\ell) \equiv (\sqrt 2 -1)\ell$.
For $t_0 \ge (2 \ell +d)/2$, all the geodesics lie entirely in a black brane background and the mutual information is given by  \cite{Tonni:2010pv}:
\bea\label{MIBTZ}
I(A,B)_{\text{thermal}}& =& \frac{1}{4 G_N}\left\{ 2 \delta {\cal L}_{\text{thermal}}(\ell) - {\text{Min}}\left[2 \delta {\cal L}_{\text{thermal}}(\ell) , \delta {\cal L}_{\text{thermal}}(2 \ell + d)+ \delta {\cal L}_{\text{thermal}}(d) \right] \right\} \nonumber \\
 &=& \left\{\begin{array}{cc}0 \,,& d \ge d_{\text{thermal}}  \\
     					  \frac c 3 \ln \left[\frac{\sinh^2 (r_H \ell /2) }{\sinh(r_H (2 \ell+d)/2) \sinh(r_H d/2)}\right] \,,& d \le d_{\text{thermal}}
					 \end{array}\right.
\eea
using \eqref{LvsellBTZren}, where 
\be
d_{\text{thermal}}(\ell) \equiv \frac{1}{r_H} \ln \left[ 1-e^{r_H \ell} + e^{2 r_H \ell} + \sqrt{(1-e^{r_H \ell} + e^{2 r_H \ell})^2 - e^{2 r_H \ell}} \right] - 2\ell
\, .
\ee
As the intervals size $\ell \to \infty$, $d_{\text{thermal}}(\ell) \to d_{ \infty} = \frac{ \ln 2}{r_H}$.

\begin{figure}[t]
\centering
\includegraphics[width=0.3\linewidth]{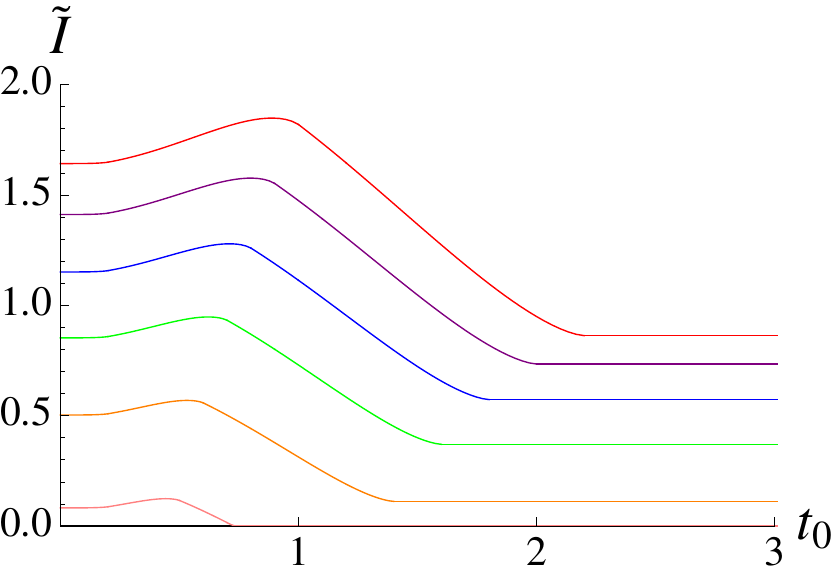}
\includegraphics[width=0.3\linewidth]{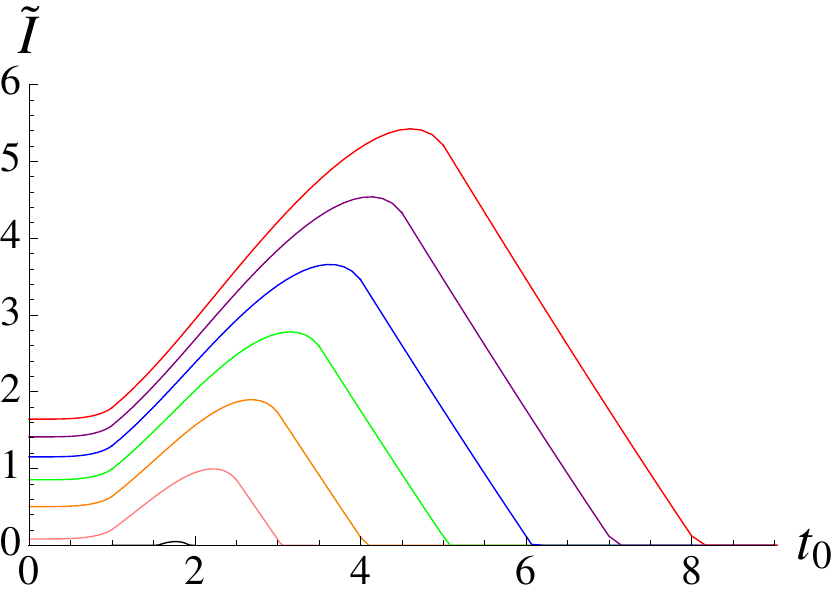}
\includegraphics[width=0.3\linewidth]{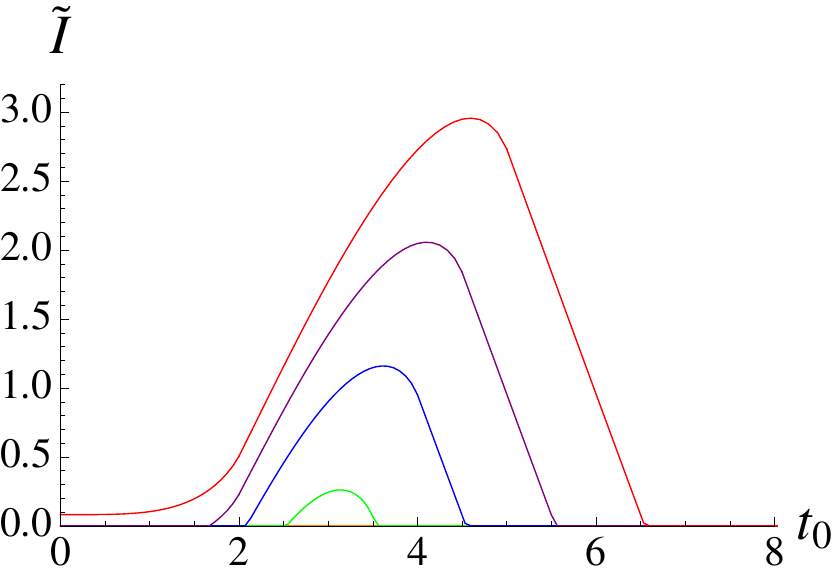}
\caption{Rescaled mutual information $\tilde I \equiv 4 G_N I$ as a function of boundary time $t_0$ for $d=0.4$ (left), $d=2$ (center) and $d=4$ (right), and $r_H =1$. The various curves correspond to different values of $\ell$ which increases from the bottom up. The left panel shows $\ell = 0.2, 0.4, \dots , 2.0$, while the center panel and the right one  show $\ell = 1,2,\dots,10$ (some of the curves in the three panels are not visible because everywhere vanishing).    
}
\label{fig:MI}
\end{figure}

From these results, we see that for a  fixed interval size $\ell$  when $d_{\text{thermal}}  \le d < d_{\text{vacuum}} $ the mutual information must vanish for $t_0 \ge (2 \ell +d)/2$.  The reason is that at these late times, all the geodesics that can potentially contribute to the mutual information (see Fig.~\ref{fig:VaidyaSpacetime}b) will lie in the black brane background.   When $d \ge d_{\text{vacuum}}$ it is zero also for $t_0 \le 0$ because at early times all the relevant geodesics lie in the empty AdS background.   When $d< d_{\text{thermal}}$  the mutual information can be non-zero for all times (Fig.~\ref{fig:dlAdS}).  

\begin{figure}[t]
\centering
\includegraphics[width=0.3\linewidth]{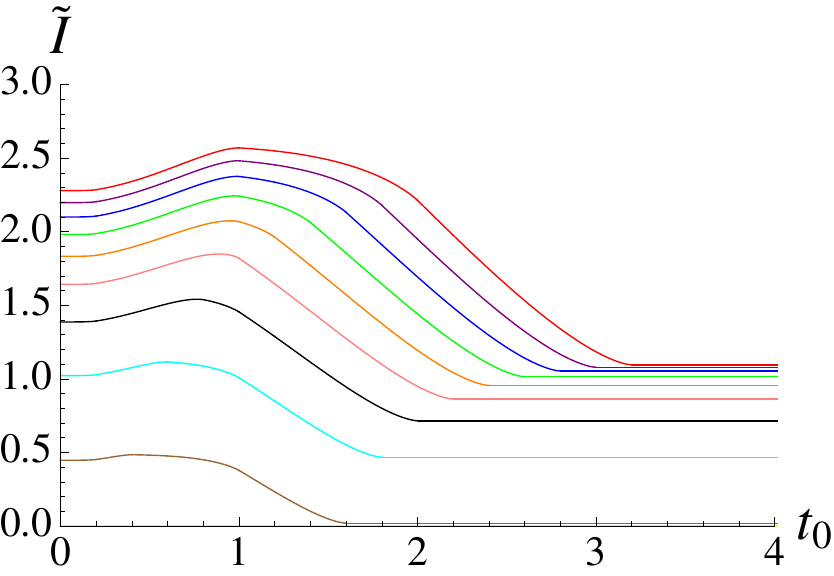}
\includegraphics[width=0.3\linewidth]{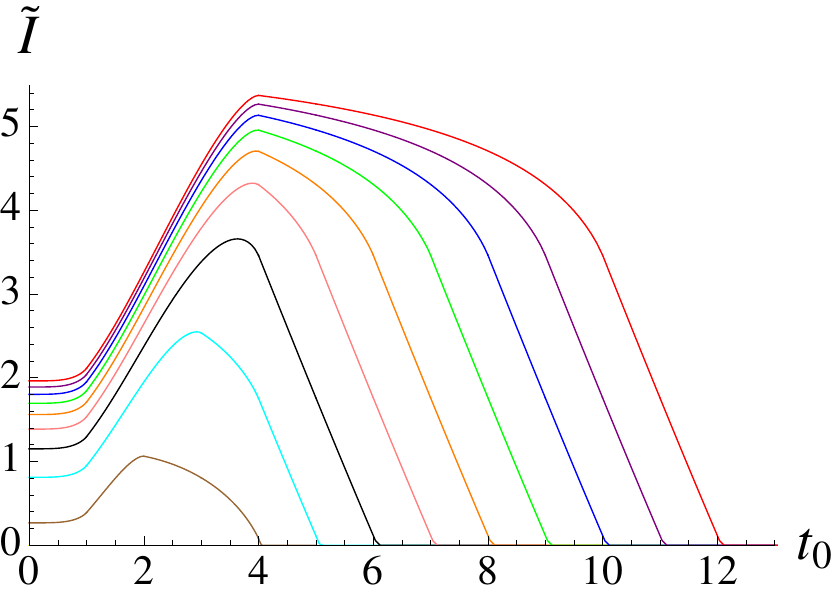}
\includegraphics[width=0.3\linewidth]{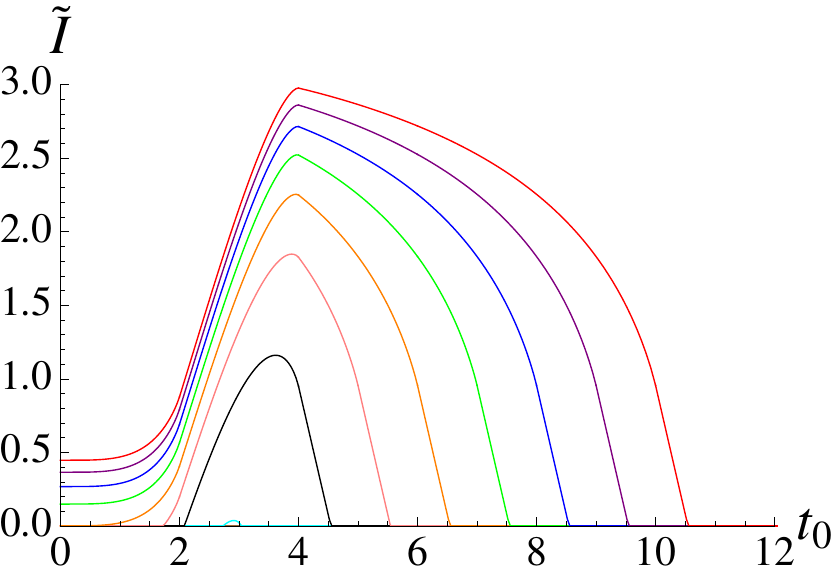}
\caption{Rescaled mutual information $\tilde I \equiv 4 G_N I$ as a function of boundary time $t_0$ for $d=0.4$, $\ell_1 = 2$ (left),  $d=2$, $\ell_1 = 8$ (center) and $d=4$, $\ell_1 =8$ (right), and $r_H =1$. The various curves correspond to different values of $\ell_2$ which increases from the bottom up. The left panel shows $\ell_2 = 0.4, 0.8, \dots, 4.0$ while the center and the right panels have $\ell_2 =  2,4, \dots, 20$. }
\label{fig:MI2}
\end{figure}

In the Vaidya background the mutual information should interpolate between the vacuum and thermal results.   The interpolation occurs in the period  $0< t_0< (2 \ell +d)/2$, when one or more geodesics intersect the shell. Crossing the shell has the effect of lowering (raising) the geodesic length for a fixed boundary separation, with respect to the black brane (AdS) background. When ${\text{Max}}(\ell/2, d/2) \le t_0 <   (2 \ell +d)/2$ only the outermost geodesic with boundary separation $2 \ell + d$ crosses the shell.  When ${\text{Min}}(\ell/2, d/2) \le t_0 <   {\text{Max}}(\ell/2, d/2)$  geodesics  with boundary separation $2 \ell +d$ and ${\text{Max}}(\ell, d)$  cross it, and so on. The mutual information in this range of times can be computed using the appropriate geodesic lengths $\delta {\cal L}_{\text{shell}}$ in \eqref{Lren} and $\delta {\cal L}_{\text{thermal}}$ in \eqref{LvsellBTZren}.  The mutual information computed in this way starts at the vacuum value at early times and ends at the thermal value at late times. Remarkably, it increases sharply to a peak at intermediate times (Fig.~\ref{fig:MI}).   These results are easily extended to a situation where the two intervals have different lengths $\ell_1$ and $\ell_2$.  The logic is identical, so we merely plot the results in Fig.~\ref{fig:MI2}.

\section{A qualitative explanation for the mutual information?}
\label{Interpretation}

Why does the mutual information increase and then decrease again in our setup?    For understanding the entanglement entropy of a single interval in the same setting it was useful to compare the AdS results with exact computations for quantum quenches \cite{AbajoArrastia:2010yt, Calabrese:2005in}.    The entanglement entropy for multiple intervals after a quench was also computed by Calabrese and Cardy \cite{Calabrese:2005in} for times and spatial scales large compared to the inverse mass gap of the early time state.   From their results we can compute the mutual information following  (\ref{MIform}).     Since the Calabrese-Cardy theory starts, unlike ours, with a gapped initial state and a type of energy deposition with short range correlations,  their results need not agree with ours \cite{AbajoArrastia:2010yt}.   However,  the explicit result  nevertheless has a sharp feature in the middle.

\begin{figure}[h]
\centering
\includegraphics[width=0.9\linewidth]{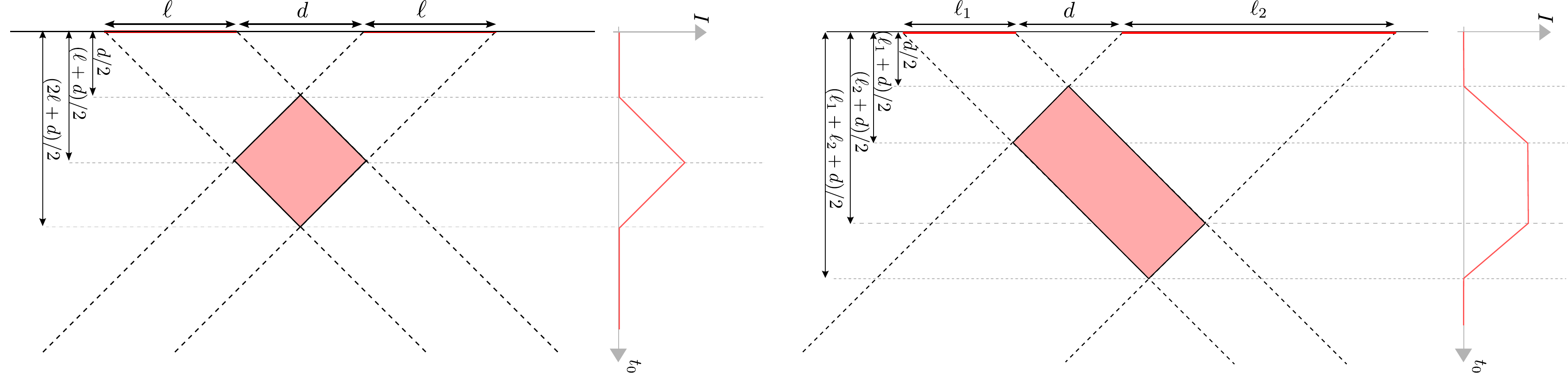}\\
({\bf a})\qquad \hfil ({\bf b}) \\
\caption{({\bf a}) A simple causality derived picture for two intervals after a quench. After the quench at $t=0$ signals propagate outwards at the speed of light in both directions. These will give correlations which contribute to the mutual information at $t=t_0$ if one signal is in each interval, which occurs only for signals originating in the shaded pink diamond. ({\bf b}) The analogous picture for two intervals of different length. }
\label{fig:MIcausality}
\end{figure}

A simple toy model developed by \cite{Calabrese:2005in} gives a physical interpretation for this principal feature of the mutual information in the quantum quench.     The idea was that before the quench there is a mass gap and a finite correlation length.  Thus, after the quench only excitations originating at almost the same point will be entangled.   At later times, such a pair of correlated particles will only contribute to the mutual information between two intervals, if each interval contains one of the two particles.   As shown in Fig.~\ref{fig:MIcausality} this naive model would lead to a linear rise and subsequent decline in the mutual information.    An extension of this toy model to the Vaidya setup reproduced the entanglement entropy of a single interval by including power law long range correlations in the initial state \cite{AbajoArrastia:2010yt}.  

To try to reproduce our results more accurately, further modifications of the toy model would be needed.  Consider, for example, the non-zero final state mutual information that is present in some cases.  This can occur in a naive particle picture if the initial energy deposition had a finite spatial correlation length, and we consider particle pairs that travel in the {\it same} direction as well as in opposite directions.   In this setup, two intervals separated by a distance $d$ will always be populated at late times with particle pairs that move in the same direction and were produced at a separation slightly larger than $d$.  If the initial correlation length was bigger than $d$ a  final state mutual information can result.


\section{Tripartite information}
\label{Tripartite}

Now consider the time evolution of the tripartite information  \eqref{TInform} for three intervals on the boundary of the dynamical Vaidya background  \eqref{eq:Vaidya}.   From \eqref{TInform}, in addition to the computations that we have already done, we need to compute $S(A\cup B \cup C)$.  Following the previous sections, we will assume that this is determined by the length of the shortest collection of geodesics connecting the endpoints of $A$,  $B$ and $C$. 

We first consider the case of three adjacent intervals $A$, $B$ and $C$, of length $\ell$, $d$ and $\ell$ respectively,  on the boundary of the Vaidya spacetime. This corresponds to  the setting considered  in Sec.~\ref{sec:MI} for the mutual information, now promoting the region of length $d$, which separates the two intervals of length $\ell$, to a third interval.

\begin{figure}[t]
\centering
\includegraphics[width=0.3\linewidth]{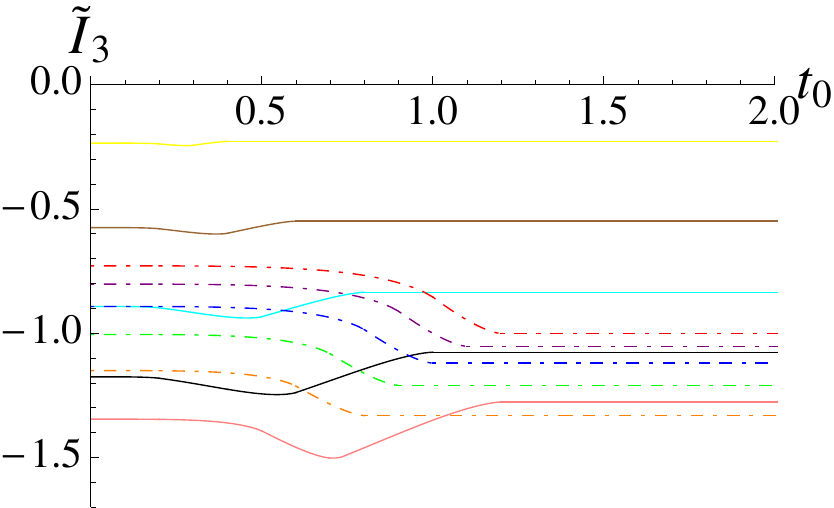}
\includegraphics[width=0.3\linewidth]{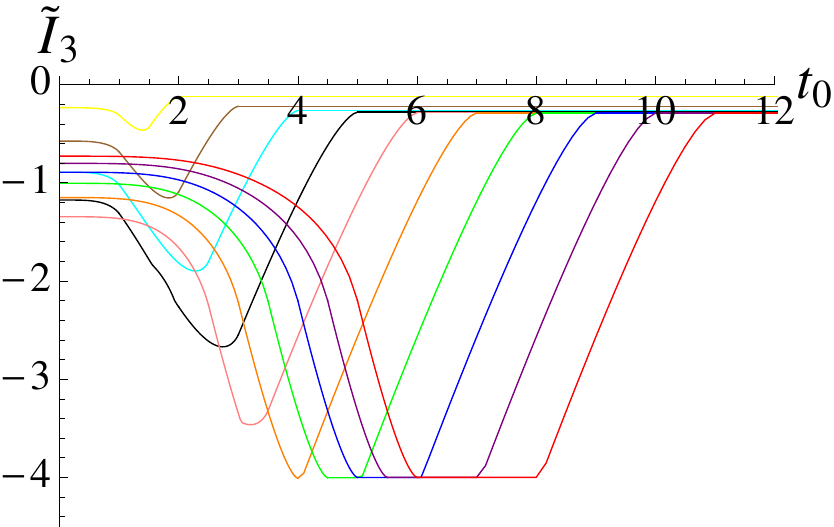}
\includegraphics[width=0.3\linewidth]{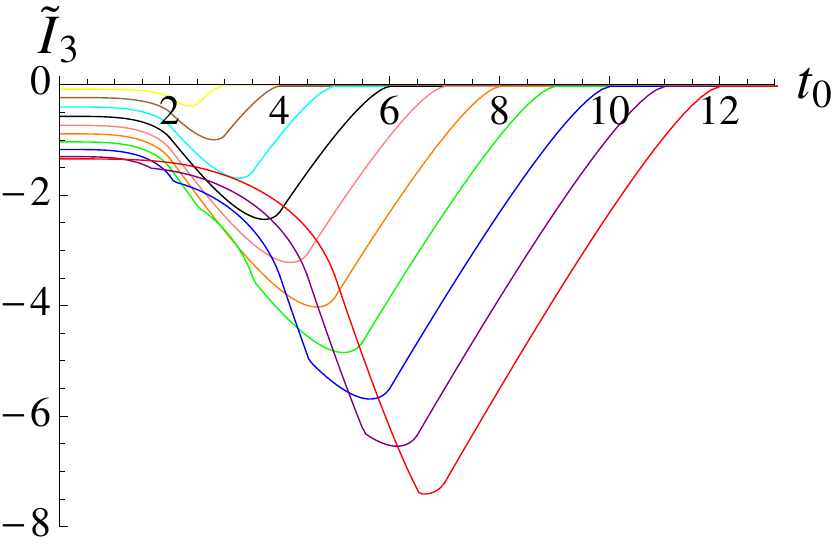}
\caption{Rescaled tripartite information $\tilde I_3 \equiv 4 G_N I_3$ for three adjacent intervals of length $\ell$, $d$ and $\ell$ as a function of boundary time $t_0$, for $d=0.4$ (left), $d=2$ (center) and $d=4$ (right), and $r_H =1$. The various curves correspond to different values of $\ell$. The left panel shows  $\ell = 0.2, \dots 1,0$ in continuous lines with $\ell$ decreasing from the bottom up, while the curves in dot-dashed have $\ell =1.2, \dots, 2.0$ increasing from the bottom up. The center and right panels have $\ell = 1,2, \dots, 10$ decreasing from right to left.}
\label{fig:TIadj}
\end{figure}

We plot in Fig.~\ref{fig:TIadj} the results for the tripartite information \eqref{TInform} for different values of $\ell$ and $d$ as a function of the boundary time $t_0$. 
For  large  values of $d$ (center and right panels of Fig.~\ref{fig:TIadj}) the tripartite information starts at the vacuum value and ends at the (larger) thermal value passing through an intermediate phase where it is more negative than either.  For small $d$ (left panel of Fig.~\ref{fig:TIadj}) and large enough $\ell$, the behavior is different.  In this case,  the vacuum value is less negative than the thermal one and the tripartite information just decreases non-linearly.   Incidentally, in this parameter range the thermal mutual information is non-vanishing (see Fig.~\ref{fig:MI}, and values of $\ell$ and $d$ in the blue dotted region of Fig.~\ref{fig:dlAdS}).

 For the value of $d$ in the center panel of Fig.~\ref{fig:TIadj}, for some values of $\ell$, the tripartite information goes through an intermediate phase where it is constant. This can be understood  looking at the first line of \eqref{TInform}. In this phase, all the contributions are constant except for $S(A \cup C )$ and $S(A\cup B \cup C)$, whose time-dependent contributions cancel each other in \eqref{TInform}. To see this,  first note that $S(A)$, $S(B)$, $S(C)$, $S(A \cup B)$ and $S(B \cup C)$ all correspond to geodesics extending in the thermal background and have a constant value. Meanwhile  $S(A\cup B \cup C)$  is given by the geodesic that ends on the two endpoints of  $A\cup B \cup C$, which crosses the shell. Finally, $S(A\cup C)$ is determined by the two geodesics enclosing a connected region in the bulk -- the outer geodesic of this pair is the same one that appears in  $S(A\cup B \cup C)$. 
Therefore the two contributions that cross the shell and are time-dependent cancel each other between $S(A\cup B \cup C)$ and $S(A\cup C)$, leading to a constant result for the tripartite information.  For large enough $\ell / d$, this constant value is approximately $\ell$-independent, as can be seen from \eqref{LvsellBTZren}. The same type of argument applies to late times, when all the geodesics extend in the thermal background.

As shown in Fig.~\ref{fig:TIadj},  we find a  non-positive tripartite information at all times. Even though we are dealing with a dynamical setup, this behavior matches  \cite{Hayden:2011ag}, where it has been shown that in holographic theories at equilibrium  the mutual information is extensive or superextensive.

\begin{figure}[htbp]
\begin{center}
\includegraphics[width=0.9 \textwidth]{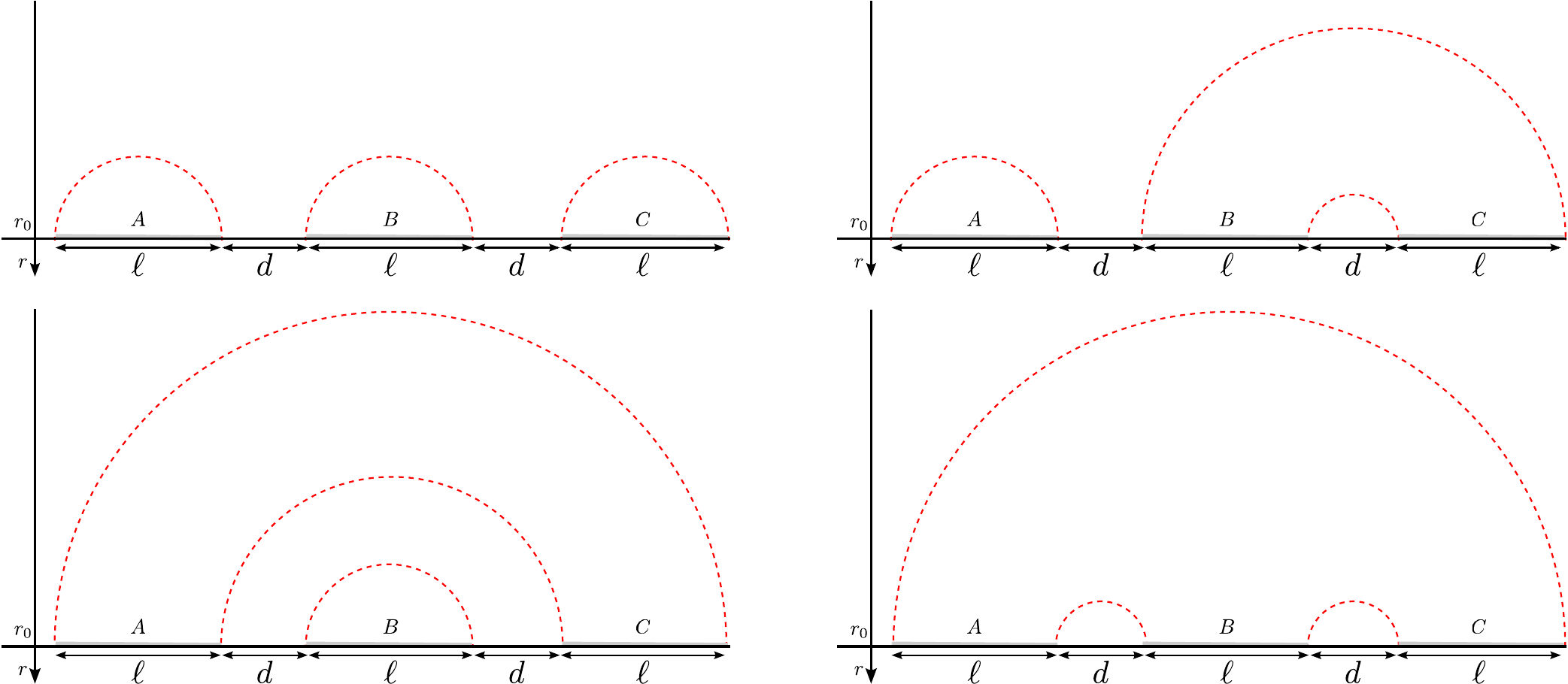}
\end{center}
\caption{A schematic representation of locally minimal surfaces for the boundary region  $A \cup B \cup C$ in  AdS$_{3}$.   The top right figure with A and C interchanged also contributes.
} \label{fig:EE3int}
\end{figure}

While the case of adjacent intervals is a direct extension of the computation of the mutual information (Sec.~\ref{sec:MI}),  the tripartite information for three disjoint intervals is more subtle. 
However, using the fact  that only non-intersecting geodesics are local minimal surfaces for two disjoint intervals, as  was the case in \eqref{mutualinfo}, the collection of geodesics that potentially contribute to the tripartite information of $A \cup B \cup C$ in the Vaidya background reduces to those shown in Fig.~\ref{fig:EE3int}.

\begin{figure}[ht]
\centering
\includegraphics[width=0.3\linewidth]{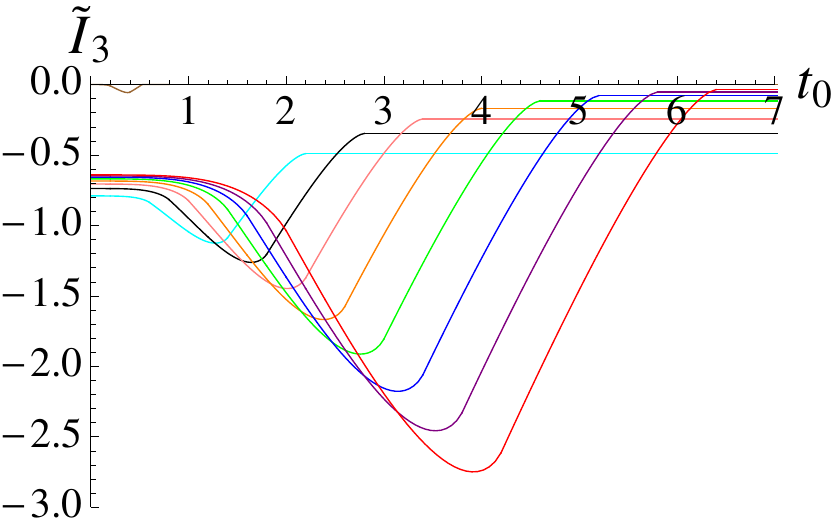}
\includegraphics[width=0.3\linewidth]{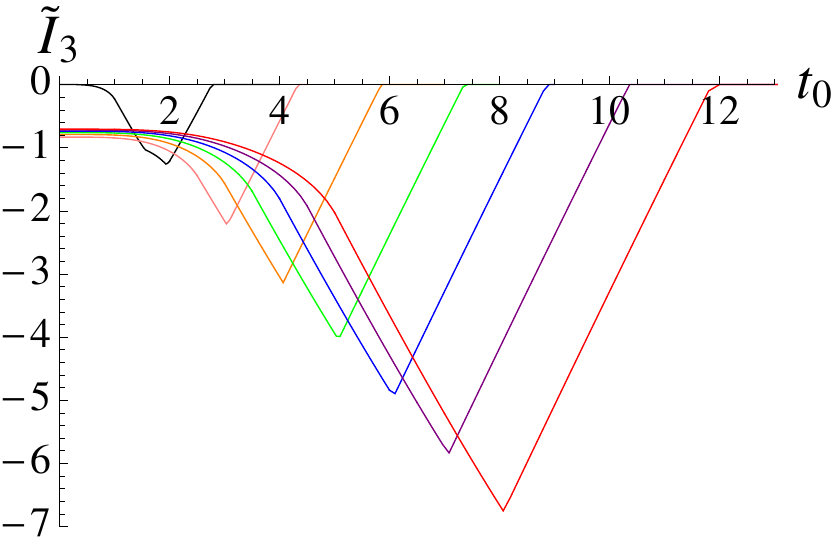}
\includegraphics[width=0.3\linewidth]{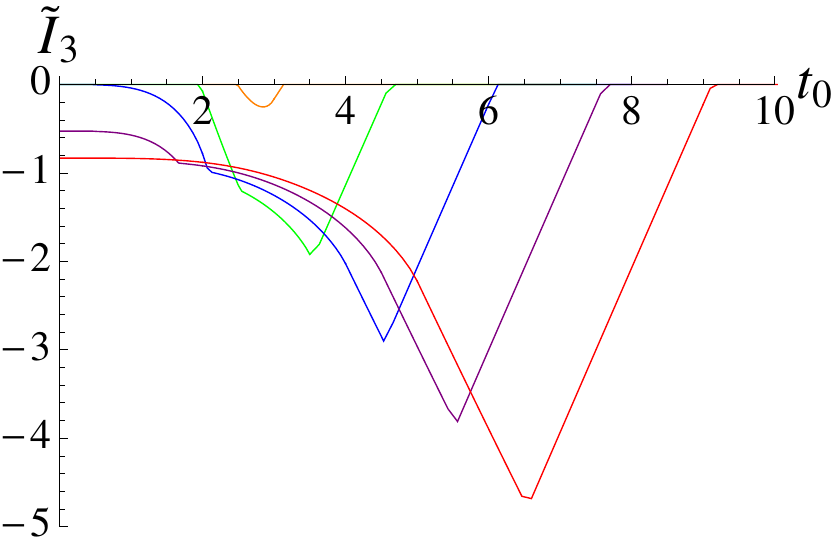}
\caption{Rescaled tripartite information $\tilde I_3 \equiv 4 G_N I_3$  as a function of boundary time $t_0$ for $d=0.4$ (left), $d=2$ (center) and $d=4$ (right), and $r_H =1$. The various curves correspond to different values of $\ell$ which decreases from right to left. The left panel shows $\ell = 0.4, 0.8, \dots , 4.0$, the center and right panels have $\ell = 1, 2, \dots, 10$ (although only $\ell = 3,\dots,10$ and $\ell = 6\dots,10$ are not everywhere vanishing in the center and right panel, respectively). }
\label{fig:TI}
\end{figure}

The time evolution of the tripartite information for three disjoint intervals of length $\ell$, each separated by a distance $d$ from adjacent ones, is plotted in Fig.~\ref{fig:TI} for various values of $d$ and $\ell$. 
It mainly shows the same features as the tripartite information of adjacent intervals.

We have found that the tripartite information computed in the Vaidya geometry is generically non-zero.  Now recall that Calabrese and Cardy found that the entanglement entropy of N disjoint intervals $(u_{2j -1}, u_{2j})$, with $1\le j \le N $ and $u_{k} < u_{k+1}$,  after a quantum quench \cite{Calabrese:2005in} evolves in time according to the formula
\be \label{CCmultiple}
S \sim 
S_\infty
+  \frac{\pi c}{12 \tau_{0} } \sum^{2N}_{k,l=1} (-1 )^{k-l-1}\text{Max}[u_{k}-t,u_{l}+t]\,. 
\ee
Here $c$ is the central charge and  $\tau_{0}$ is  a parameter of the same order as the inverse mass gap (refer to \cite{Calabrese:2005in} for more details). This formula holds only in the limit where the time $t$ and all separations $|u_{k} -u_{l}|$ are much larger than the initial inverse mass gap. Using \eqref{CCmultiple}, it follows that the time-dependent part of the  tripartite information vanishes in this regime because each contribution associated to an extremum $u_{j}$ appears an equal number of times with a minus and a plus sign in \eqref{TInform}. 
In contrast, the tripartite information for the Vaidya background  (Fig.~\ref{fig:TI}) is generally time-dependent and is only constant (and vanishing) if $I(A,B)$, $I(A,C)$ and $I(A,B\cup C)$ in \eqref{TInform} are all identically zero.

The results of Calabrese and Cardy need not agree with ours since they start with a gapped initial state which lacks long-range correlations. Nevertheless it is interesting that important qualitative features of the time evolution of the mutual information matched between their model and ours, while the behavior of the tripartite information is sharply different. In any case, the disagreement shows clearly that the simple causality argument of Sec.~\ref{Interpretation}, where interactions were not taken into account, is not enough to capture the qualitative
features of our strongly coupled two-dimensional dynamics.


\section*{Acknowledgments}

This research is supported by by DOE grant DE-FG02-95ER40893, the Belgian Federal Science Policy Office through the Interuniversity Attraction Pole IAP VI/11, by FWO-Vlaanderen through project G011410N and the National Research Foundation of Korea (NRF) grant funded by the Korea government (MEST) through the Center for Quantum Spacetime (CQUeST) of Sogang University with grant number 2005-0049409. AB and FG are Aspirant FWO. VB and BC thank the ICTP, Trieste for hospitality while this paper was being completed.


\end{document}